\documentclass[aps,prl,preprint,superscriptaddress]{revtex4}
\usepackage[dvips]{graphicx}
\usepackage{pnastwoF}
\usepackage{amssymb,amsfonts,amsmath}

\begin{document}
\title{Structural Fluctuations of Microtubule Binding Site of KIF1A in Different Nucleotide States}

\author{Ryo Kanada}
\email[]{kanada@theory.biophys.kyoto-u.ac.jp}
\affiliation{Department of biophysics, Kyoto University, Kyoto 606-8502, Japan}
\affiliation{Cybermedia Center, Osaka University, Toyonaka 560-0043, Japan}
\author{Fumiko Takagi}
\affiliation{Formation of Soft Nanomachines, Core Research for Evolutional Science and Technology, Japan Science and Technology Agency, Osaka, Japan}
\affiliation{Cybermedia Center, Osaka University, Toyonaka 560-0043, Japan}
\author{Macoto Kikuchi}
\affiliation{Cybermedia Center, Osaka University, Toyonaka 560-0043, Japan}
\affiliation{Formation of Soft Nanomachines, Core Research for Evolutional Science and Technology, Japan Science and Technology Agency, Osaka, Japan}
\affiliation{Graduate School of Frontier Biosciences, Osaka University, Osaka 565-0871, Japan}
\affiliation{Department of Physics, Osaka University, Toyonaka, Osaka 560-0043}

\begin{abstract}
How molecular motors like Kinesin regulates the affinity to the rail protein in the process of ATP hydrolysis (ATP $\to$ ADP${\bf \cdot}$Pi $\to$ ADP + Pi) remains to be uncovered. 
To understand the regulation mechanism, we investigate the structural fluctuation of KIF1A in different nucleotide states that are realized in the ATP hydrolysis process by molecular dynamics simulations of G\={o}-like model.
We found that $\bf{\alpha 4}$ helix, which is a part of the microtubule (MT) binding site, changes its fluctuation systematically according to the nucleotide states.
In particular, the frequency of large fluctuations of $\alpha 4$ strongly correlates with the affinity of KIF1A for microtubule.
We also show how the strength of the thermal fluctuation and the interaction with the nucleotide affect the dynamics of microtubule binding site.
These results suggest that KIF1A regulates the affinity to MT by changing the flexibility of $\alpha 4$ helix according to the nucleotide states.
\end{abstract}
\maketitle

\section{Introduction}

The linear biological molecular motor is a nano-machine that converts the chemical energy produced by the ATP hydrolysis into the mechanical work such as cellular transport and cell divisions~\cite{howard2001, alberts2002}.
In contrast to macroscopic artificial machines, biomolecular motors work under a noisy environment such as the cell.
In fact, the thermal fluctuation should be appreciable, since the free energy released by one ATP hydrolysis cycle is only $\sim 20 k_BT$. 
Despite several decades of investigations, the detailed mechanism by which motor proteins use ATP to move along the rail protein (cyto-skeletal filament) is not understood.

According to the nucleotide states that are realized in the ATP hydrolysis cycle, a linear motor generally has two binding modes to the rail protein: ``strong'' binding mode and ``weak'' binding mode~\cite{howard2001, kikkawa2004, okada2000}.
In the strong binding mode the motor attaches itself to the rail protein tightly, whereas in the weak binding mode the affinity of motor for rail protein gets lower and the motor readily detaches from the rail protein.   
The mechanism of switching between these two modes, however, is still unknown. 
To reveal this regulating mechanism, we focus on KIF1A (Kinesin-3) motor considering that a lot of experimental data have been accumulated so far both on structural and biochemical properties~\cite{kikkawa2004, okada2000, okada1999, okada2003, kikkawa2001}.

KIF1A is a single-headed molecular motor and can move processively and unidirectionally along a microtubule (MT) by using ATP hydrolysis reaction~\cite{okada2000}. 
The recent experiments which used several different nucleotide analogs (ATP analog, ADP$\cdot$Pi analog, etc.) have revealed the structure of KIF1A and the equilibrium dissociation constant for MT ($K_d$) in each nucleotide states~\cite{kikkawa2004, okada2000, okada1999, okada2003, kikkawa2001, kikkawa2000}.
It was found that during the ATP hydrolysis process KIF1A switches its binding modes for MT as follows: in the nucleotide free state and the ATP bound states, the  ``strong'' binding mode is realized in which KIF1A attached tightly to the specific site of MT.
In ADP$\cdot$Pi state which is realized as a result of ATP hydrolysis reaction (ATP$\to$ADP$\cdot$Pi) on head, the binding modes of KIF1A for MT varies according to the relative configuration of ADP and $\gamma$-Pi.
Just after the hydrolysis, in ``the early ADP-phosphate state'' where the distance between ADP and $\gamma$-Pi is small (about $3\,\rm\AA$), the binding mode is still ``strong''.
On the other hand, just before the phosphate ($\gamma$-Pi) release from the KIF1A head, in ``the late ADP-phosphate state'' where the distance between ADP and $\gamma$-Pi is more than $10\,\rm\AA$, KIF1A realizes the ``weak'' binding modes in which the affinity of KIF1A for MT becomes much lower than the  ``strong'' binding modes.
After the phosphate release, the ADP bound states realizes the ``weak'' binding mode.
Here, we have the following question: ``Which part of KIF1A regulates the switching of the affinity for MT according to the nucleotide states and what is its mechanism?''

A recent experiment which includes the structural analysis suggested that mainly two loops (L11, L12) play an important role in controlling the binding modes of KIF1A for MT \cite{kikkawa2004}.
But it is also reported that other than two loops, there are a number of interaction sites with MT such as $\beta 5 a$, L8, $\beta 5 b$, $\beta 7 a$, $\beta 7 b$, $\alpha 4$, $\alpha 5$, which includes switch II region~\cite{kikkawa2000}. Then we have another question: "other than two loops (L11, L12), are there any part to regulate the strength of binding on MT?"

For conventional kinesin which belongs to the same superfamily as KIF1A, it was reported that a large B-factor of the X-ray crystal structure does not correspond to a functionally important fluctuation~\cite{sack1999}. 
Also, the normal mode analysis based on the elastic-network model has revealed that small fluctuations in the elastic regime are insufficient to capture the conformational change of kinesins, including KIF1A~\cite{zheng2003}. 
Furthermore, by a simulation study of a realistic lattice model it was reported that the conventional kinesin in ADP state (PDB ID code: 1BG2) exhibits partial folding/unfolding at the functionally important regions including the microtubule binding sites (switch II region) other than two loops (L11, L12)~\cite{kenzaki2007}.
These results indicate that larger-scale structural fluctuations beyond the elastic regime are essential to the function of Kinesin.

The energy landscape theory of protein folding has been accepted widely in the last decade. The theory states that proteins have a funnel-like energy landscape toward the native structure~\cite{brygenlson1995, onuchic1997}.
The G\={o}-like model is certainly the simplest class of model that realizes a funnel-like landscape \cite{go1983} and has successfully described the folding process of small proteins.
Recently, some attempts have been made to simulate a larger structural change beyond the elastic regime by G\={o}-like model ~\cite{koga2006}.
For conventional kinesin, in particular, G\={o}-like model simulation succeeded in revealing the important function, such as the mechanism how the internal strain regulates the fluctuation of nucleotide binding site and how the directional stepping is controlled ~\cite{hyeon2007, hyeon2007_2}.
To analyze the large structural fluctuations which is relevant to the mechanism of MT binding, we use one of the standard Go-models~\cite{clementi2000}.

In this article, we investigate the structural fluctuations of KIF1A in various nucleotide states at thermal equilibrium by means of the molecular dynamics simulation using a simple G\={o}-like model~\cite{clementi2000} in order to reveal which part of KIF1A plays an important role in switching the binding strength to MT. 
We also make simulations that includes nucleotide molecule explicitly to investigate the effect of presence of the nucleotide following the recent work on myosin by Takagi and Kikuchi~\cite{takagi2007}

\section{Results}

\subsection{Reference structure}
We focus on the structures of KIF1A in the following five intermediate states in the ATP hydrolysis: 1)  the early ATP bound state in which the nucleotide binding pocket is still open. We call it as ``ATP 1 state''. 2) the late ATP bound state, or, pre-hydrolysis state, in which the nucleotide binding pocket is closed. We call it as ``ATP 2 state''. 3) the early ADP-phosphate state, or, post-hydrolysis state.  We call it as ``ADP$\cdot$Pi 1 state''. 4) the late ADP-phosphate state, or, pre Pi release state. We call it as ``ADP$\cdot$Pi 2 state.\ 5) ``ADP state''.
The equilibrium dissociation constants $K_d$ from MT in 1)-3) states are small (strong binding mode), whereas ones in 4)-5) states is large (weak binding mode)~\cite{kikkawa2004, okada2000}.

We construct G\={o}-like models for each of the five states.
For the reference structures of the ``native'' states that is required to define G\={o}-like model, we employ X-ray structures of KIF1A with different nucleotide analogs.
As ``ATP 1 state'' and ``ATP 2 state'', we use 1I6I~\cite{kikkawa2001} and 1VFV~\cite{kikkawa2004} (PDB ID codes) structures that are realized in AMP-PCP and AMP-PNP (ATP analog) bound state, respectively. 
As ``ADP$\cdot$Pi 1 state'' and ``ADP$\cdot$Pi 2 state'', we use 1VFX \cite{kikkawa2004} and 1VFZ \cite{kikkawa2004} (PDB ID codes) structure which are realized in ADP-ALFx  and ADP-Vi (ADP$\cdot$Pi analog) bound state, respectively.
Finally, as ``ADP state'', we use 1I5S \cite{kikkawa2001} (PDB ID codes) structure which is realized in ADP nucleotide bound state.

For each of the five intermediate states, we made simulations of about $5.0\times 10^7 \sim 1.0 \times 10^8$ steps.
The temperature was set lower than the folding temperature which was estimated by MD simulations ($T_{f} \sim 1.4-1.45$ for all the models).
We mainly show the results at $T= 1.3$ in the following sections.
Simulations were made by solving Langevin equation.
The detail of the simulation is described in Model and Method section.

\subsection{Distance root-mean-square deviation for all C$_{\alpha}$ pair (dRMSD)}

The Distance root-mean-square deviation  (${\rm dRMSD}_{ij}$) between i-j C$_{\alpha}$ pair relative to the native-structure is defined as 
\begin{align} 
{\rm dRMSD}_{ij} = \sqrt{\frac{1}{N_s} \sum_{step=1}^{N_s}
(r_{ij}(step)-<r_{ij}>)^{2}}, \label{eq:def_dRMSD}
\end{align}
where $N_s$ is the step number of simulation and $<r_{ij}>$ is the time average of the distance between i-j pair. 

\begin{figure}[t]
	\begin{center}
	\includegraphics[width= 7.0cm]{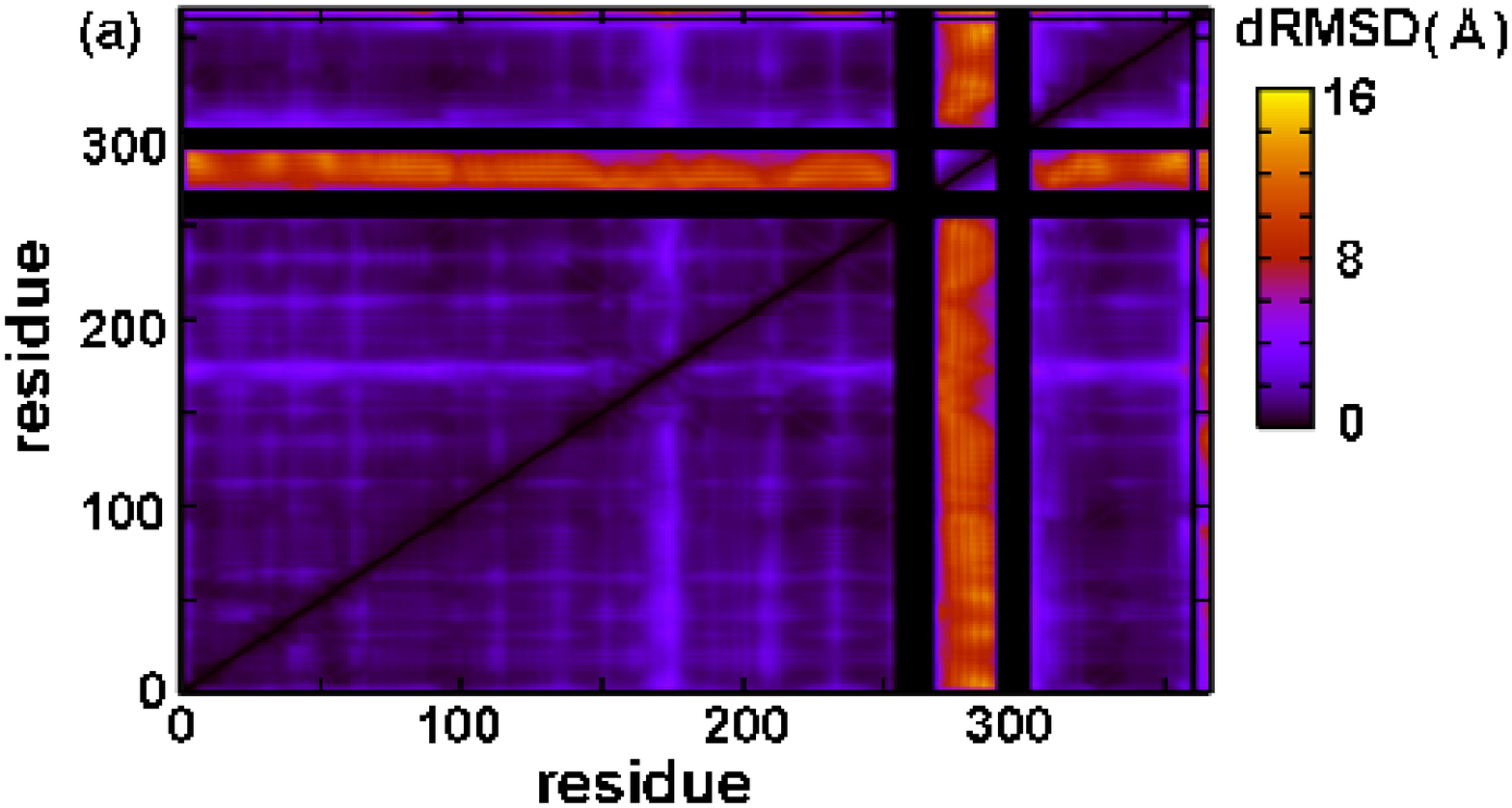}
	\includegraphics[width = 7.0cm]{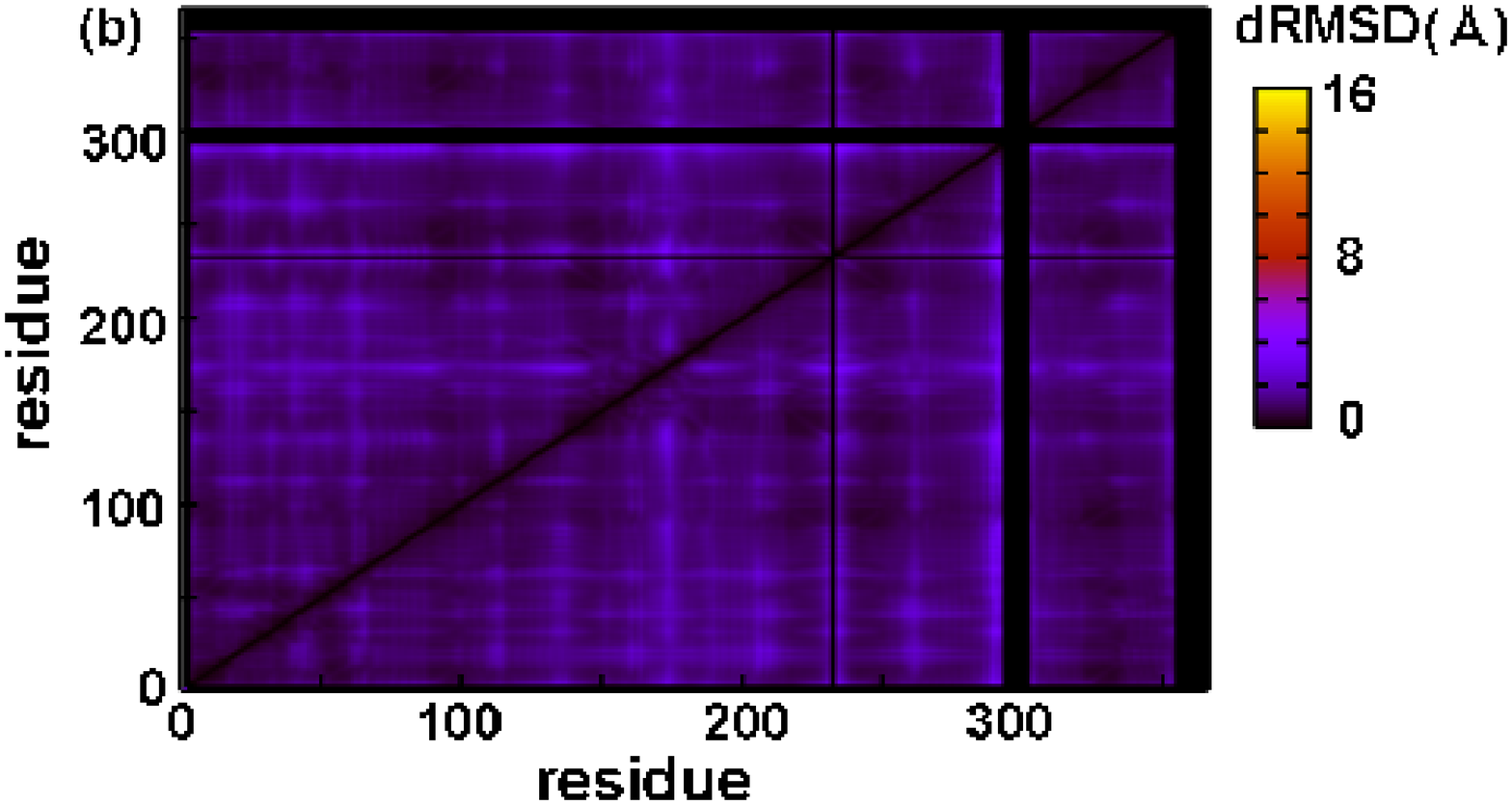}
	\end{center}
	\vspace{-2mm}
	\caption{
The simulated result of dRMSD contour plots for all C$_{\alpha}$ pair. 
(a) the case in ``ATP 1'' (AMP-PCP bound) state which is a representative of strong binding states.
(b) the case in ``ADP$\cdot$Pi 2'' (ADP$\cdot$Pi bound) state which is a representative of weak binding states.
\label{fig:jfig_contour}}
\end{figure}

In the five intermediate states, we investigated the dRMSD for all C$_{\alpha}$ pair. 
We found a qualitative difference for the dRMSD between the strong binding modes (``ATP 1'', ``ATP 2'', ``ADP$\cdot$Pi 1'') and the weak binding modes (``ADP$\cdot$Pi 2'', ``ADP'').
Here we show two contour plots of dRMSD: Fig.~\ref{fig:jfig_contour} (a) is for ``ATP 1 state'', which is a representative of strong binding states.  
(b) is for ``ADP$\cdot$Pi 2 state'', which is a representative of the weak binding states.
We omitted dRMSD for the gap regions (missing residues) which includes loop L11 (residues around 260-270) and L12 (residues around 290-300) from Fig.~\ref{fig:jfig_contour}.
It can be seen in the figure that in the strong binding state (``ATP 1'') a  large structural fluctuations are localized at 270-290 residue (dRMSD $\sim 10\,\rm\AA$).
On the other hand, in the weak binding state (``ADP$\cdot$Pi 2''), no such large fluctuation is appreciable.
The part (residue number: 270-290) that exhibits large fluctuations in the strong binding state corresponds to $\alpha_{4}$ helix, which is  a candidate of MT binding sites~\cite{kikkawa2000}. 
This result suggests that the fluctuations of $\alpha_4$ helix correlates to the strength of MT binding.
To confirm it, next we investigate the dynamics of $\alpha_{4}$ helix.\\

\subsection{Dynamics of $\alpha_4$ helix}
We calculated two quantities: 
$\Delta \Theta_{\alpha4}$ is the angle between $\alpha 4$ helix (residues 270-290) and $\alpha 3$ helix (residues 170-190), which we choose as the representative of the head since the structure of $\alpha 3$ helix is considerably stable and attached to the other part of the head tightly. 
The orientation of two helices is presented in Fig.~\ref{fig:jfig_snapshot} (a), which is a snapshot of the typical simulated structure of KIF1A in ``ATP 1'' state. 
$\Delta \Theta_{\alpha4}$ is the difference between the angle of the native structure and one of given conformation and is defined as 
\begin{align}
\Delta \Theta_{\alpha4} = \frac{180}{\pi} \Bigl[ \cos^{-1}
\left(\frac{
({\bf r}_{\alpha 3} \cdot {\bf r}_{\alpha 4})
}
{{r_{\alpha 3}}{r_{\alpha 4}}}
\right)
-\cos^{-1}
\left(\frac{
({\bf r}^{(0)}_{\alpha 3} \cdot {\bf r}^{(0)}_{\alpha 4})
}
{{r^{(0)}_{\alpha 3}}{r^{(0)}_{\alpha 4}}}
\right)
\Bigr] \label{eq:def_delt_theta_alph4}
\end{align}
where the vectors ${\bf r}_{\alpha 3}, {\bf r}_{\alpha4}$ are given by ${\bf r}_{\alpha 3} = {\bf r}_{190}-{\bf r}_{170}$, ${\bf r}_{\alpha 4} = {\bf r}_{290}-{\bf r}_{270}$ respectively.

The other quantity is the order parameter $Q_{\alpha 4}$, which is the native contact fraction between $\alpha 4$ helix (residue 270-290) and the other parts formed in a given conformation.

\begin{figure}[t]
	\begin{center}
	\includegraphics[width=7.0cm]{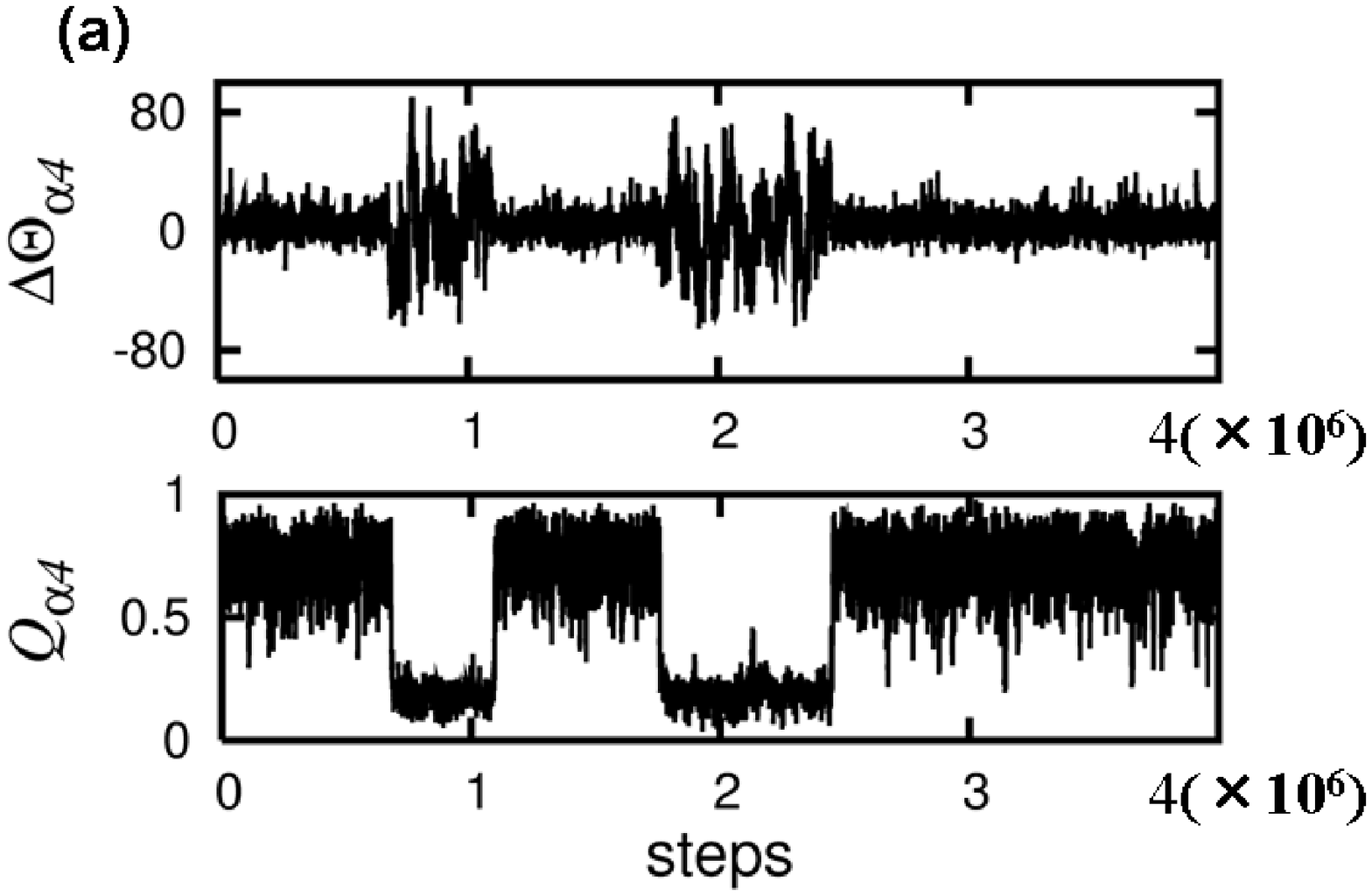}
	\includegraphics[width = 7.0cm]{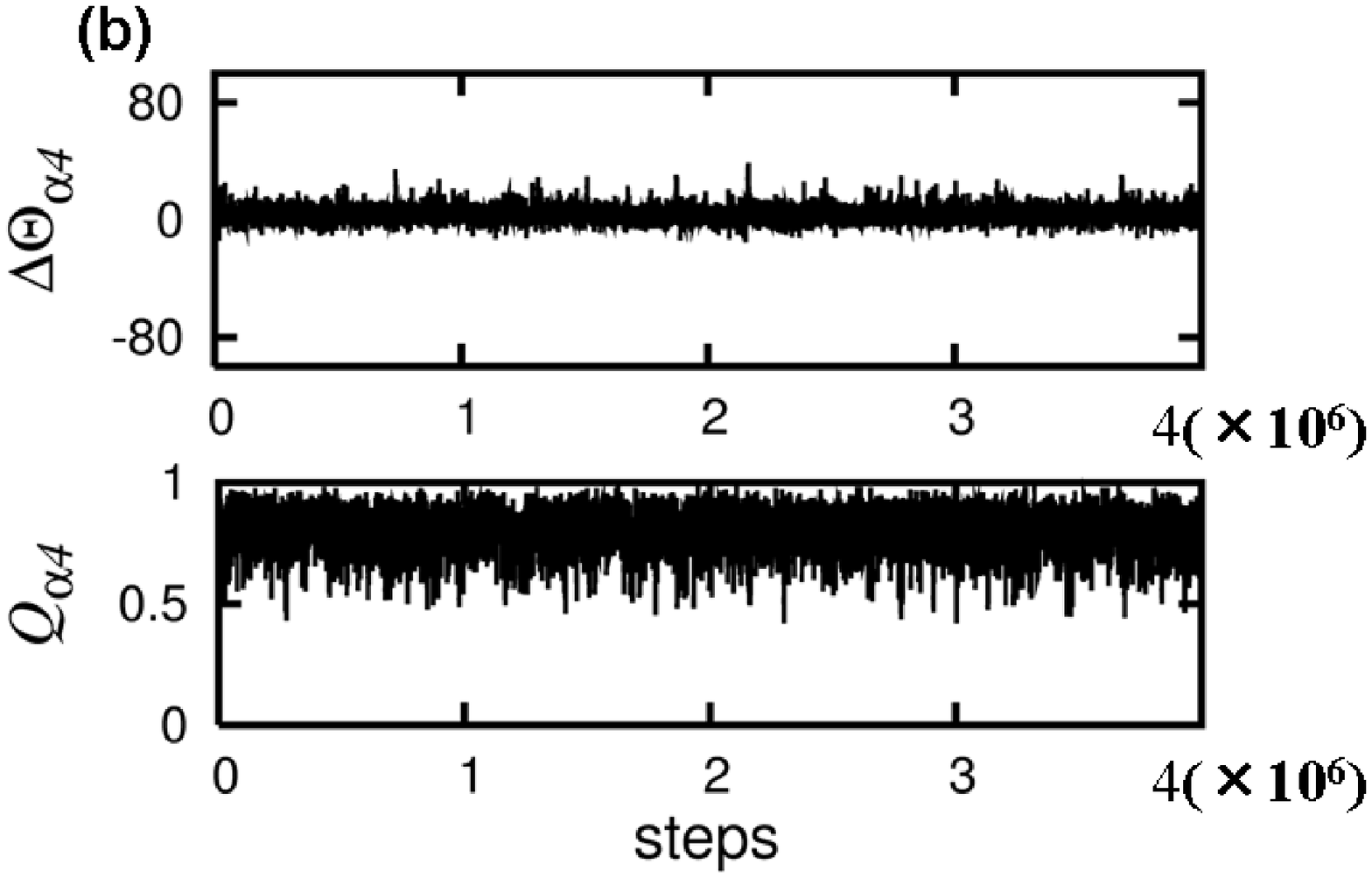}
	\end{center}
	\vspace{-2mm}
	\caption{
A typical time courses of $\alpha 4$ dynamics in strong binding state ``ATP 1 state'' (a) and weak binding state ``ADP$\cdot$Pi 2 state'' (b).
The upper figures show the time dependence of the value $\Delta \Theta_{\alpha 4}$ (Eq.~\ref{eq:def_delt_theta_alph4}) which is the angle between $\alpha 4$ helix and KIF1A catalytic core ($\alpha 3$ helix).
The lower figures are the time dependence of the measure $Q_{\alpha 4}$ which is the fraction of native contacts between $\alpha 4$ and the other parts of head.
\label{fig:jfig_traject}}
\end{figure}

\begin{figure}[t]
	\begin{center}
	\includegraphics[width=3.5cm,height=4.5cm]{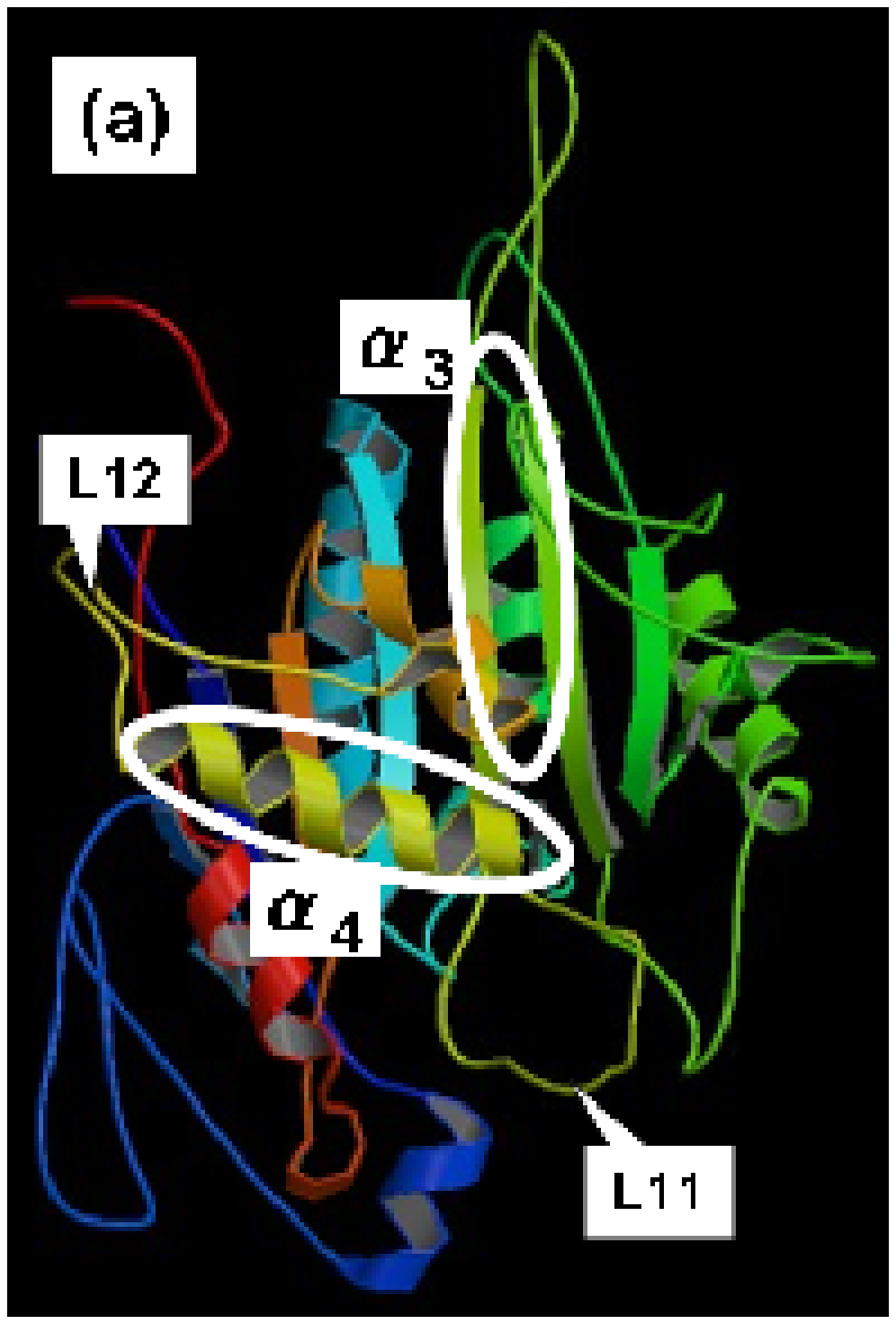}
	\includegraphics[width = 3.5cm]{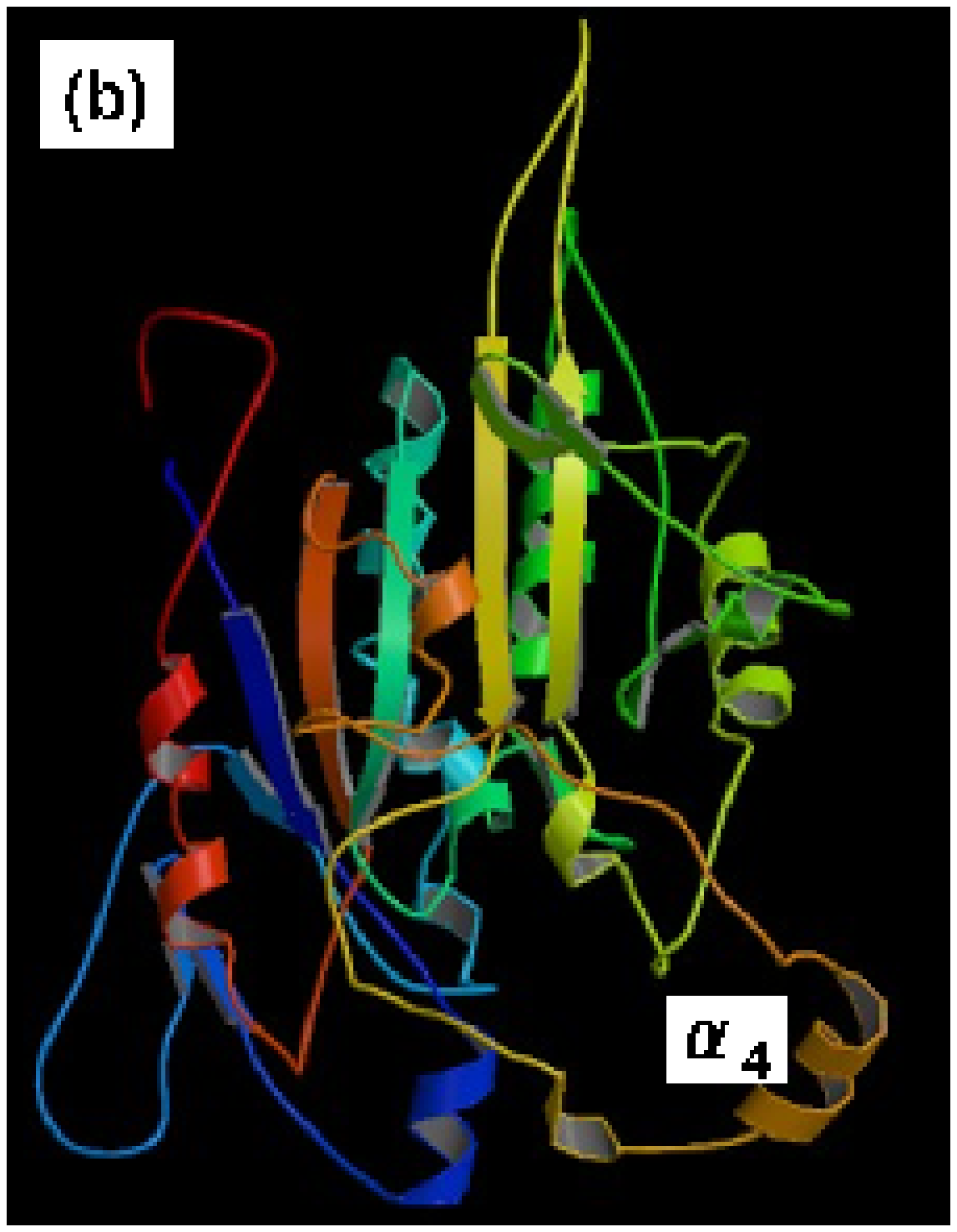}
	\end{center}
	\vspace{-2mm}
	\caption{
The typical simulated structures of KIF1A in ``ATP 1'' (AMP-PCP bound) state. 
Fig.(a) is the snapshot when the native contacts between $\alpha 4$ and the other part are almost formed ($Q_{\alpha 4} \sim 0.75$).
Fig.(b) is the snapshot when the burst (large fluctuation) occurs (on $\alpha 4$) ($| \Delta \Theta_{\alpha 4}| \sim 80^{\circ}$).
\label{fig:jfig_snapshot}}
\end{figure}

The typical time course of these two values are shown in Fig.\ref{fig:jfig_traject}  (a) ``ATP 1'' (AMP-PCP bound) state and (b) ``ADP$\cdot$Pi 2'' (ADP-Vi bound) state.
It can be seen in the figure that $\alpha 4$ helix in the strong binding state ``ATP 1'' exhibits considerably large fluctuations in the angle ($|\Delta \Theta_{\alpha 4}| \sim 80^{\circ}$) intermittently, whereas in the weak binding state ``ADP$\cdot$Pi 2'' the fluctuation of the angle is kept small ($|\Delta \Theta_{\alpha 4}| < 15^{\circ}$). 
We call the intermittent large fluctuations of $\alpha 4$ helix as ``burst''.
The burst actually is a partial unfolding of the helix. 
We also find that the burst is accompanied by decrease of the order parameter $Q_{\alpha 4}$ ($\sim 0.15$).
The native contact fraction of the whole head $Q_{\rm total}$ hardly change (at most $10\,\%$ decrease) at the burst (time course is not shown).
These facts suggest that breaking the contacts between $\alpha 4$ and the other parts of head by the thermal fluctuation induces the burst at $\alpha 4$.
The snapshot at the burst is shown in Fig.~\ref{fig:jfig_snapshot}(b), which corresponds to $|\Delta \Theta_{\alpha 4}| \sim 80^{\circ}$ and $Q_{\alpha 4} \sim 0.15$).
We find that a folded part of $\alpha 4$ is located distant from the other parts of the head.  

We also made similar simulations for other nucleotide states (``ATP 2'', ``ADP$\cdot$Pi 1'', and ``ADP''), and observed the burst of $\alpha 4$ helix.
Then, we investigate the nucleotide state dependence of the burst frequency by making histogram of the contact fraction $Q_{\alpha 4}$ from the simulated time course.\\

\subsection{Nucleotide state dependence of the frequency for burst fluctuation of $\alpha 4$ helix}
\begin{figure}[t]
	\begin{center}
	\includegraphics[width=7.0cm]{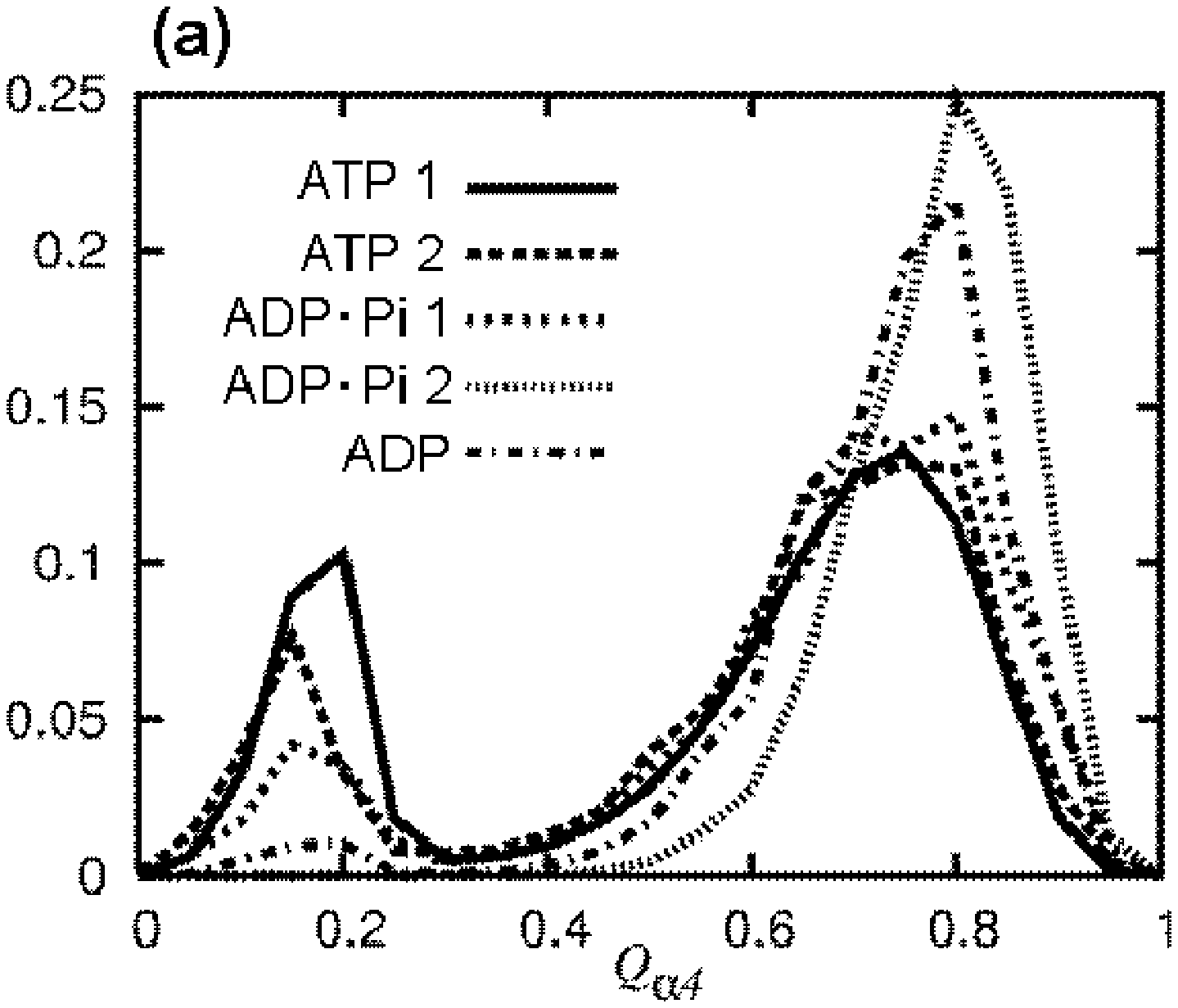}
	\includegraphics[width = 8.5cm]{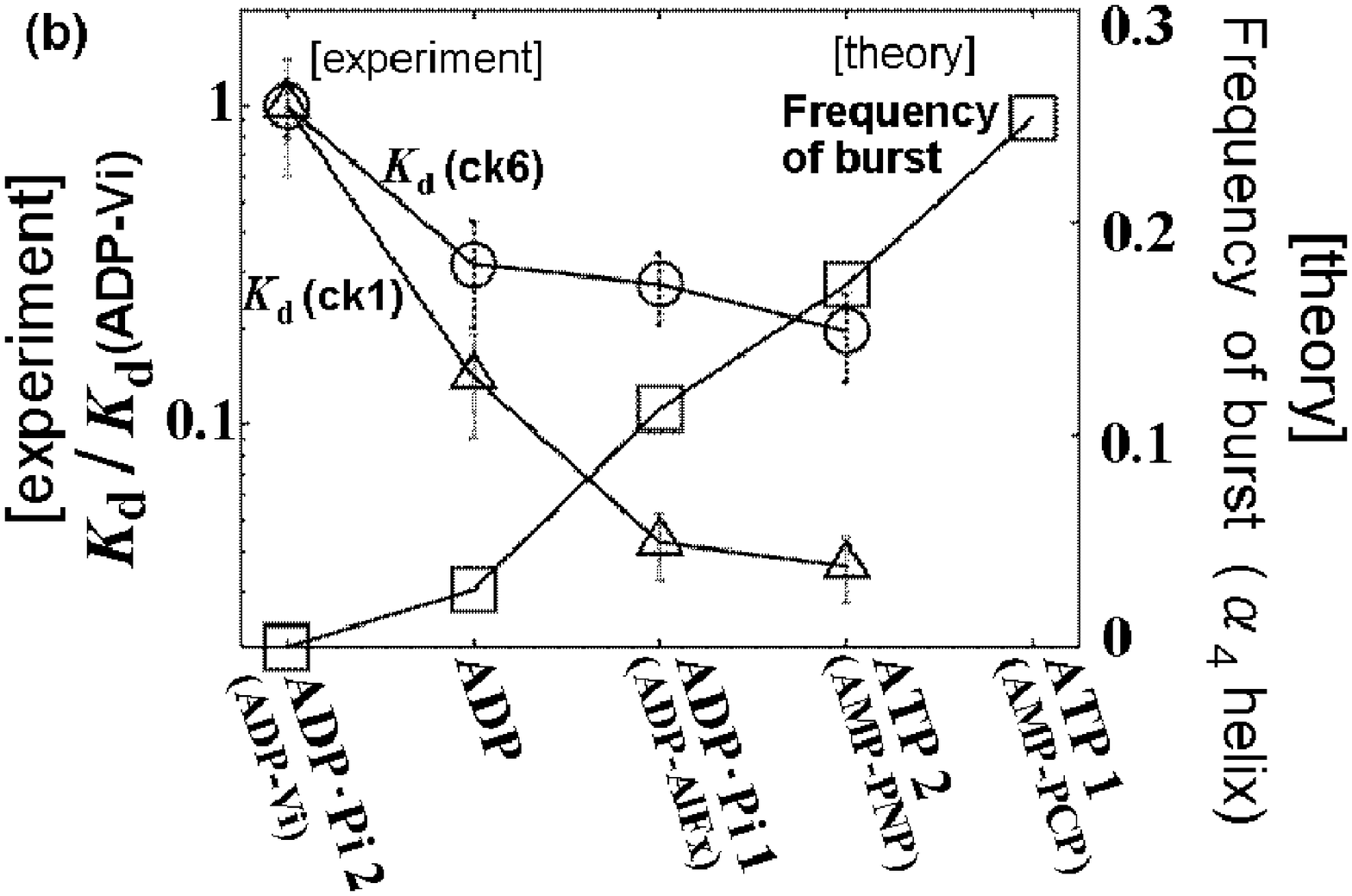}
	\end{center}
	\vspace{-2mm}
	\caption{
Fig. (a) is the histogram of the contact fraction $Q_{\alpha 4}$ in five intermediate states.
Fig. (b) shows the nucleotide state dependence of the frequency of burst at temperature $T = 1.3$ (theory) and the equilibrium dissociation constants $K_{d}$ which is normalized by the one for ADP-Vi state $K_d {\rm (ADP-Vi)}$. 
\label{fig:jfig_histo_kd}}
\end{figure}

The histograms of contact fraction $Q_{\alpha 4}$ for different nucleotide states are shown in Fig.~\ref{fig:jfig_histo_kd} (a).
The distribution of $Q_{\alpha 4}$ varies according to the nucleotide state, while the peak position stays around $\sim 0.8$ and $\sim 0.2$ irrespective to the nucleotide states.
The ratio of the lower peak ($Q_{\alpha 4} < 0.4$) corresponds to the frequency of the burst in each state. 
The nucleotide state dependence of the frequency of the burst is shown in Fig.\ref{fig:jfig_histo_kd} (b).
The horizontal axis represents the nucleotide states, which is arranged in the frequent order of the burst.
The order 1) ``ATP 1'', 2) ``ATP 2'', 3) ``ADP$\cdot$Pi 1'', 4) ``ADP'', and 5) ``ADP$\cdot$Pi 2'' roughly corresponds to the reaction sequence that occurs on the KIF1A head: ``ATP 1'', 2) ``ATP 2'', 3) ``ADP$\cdot$Pi 1'', 4) ``ADP$\cdot$Pi 2'', and 5) ``ADP''~\cite{kikkawa2004}.

From Fig.~\ref{fig:jfig_histo_kd} (b), we find that the frequency of the burst obtained by the MD correlates negatively with the experimental value of the equilibrium dissociation constants $K_{d}$ \cite{okada2000, kikkawa2004} both of CK6 (wild type KIF1A) and CK1 (a mutant KIF1A that lacks the part of L12 which has positive charge and interacts with MT in the weak binding mode).
Since the equilibrium dissociation constant is an inverse of the binding strength of KIF1A for the microtubule, the frequency of the burst of $\alpha 4$ correlates positively to the binding strength for microtubule.
These results indicate that $\alpha 4$ helix, which is a part of MT binding sites, controls the affinity for microtubule by changing its flexibility according to the nucleotide state.

\begin{figure}[t]
	\begin{center}
	\includegraphics[width = 7.0cm]{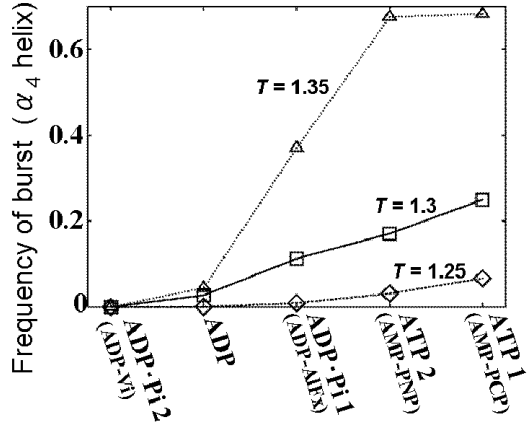}
	\end{center}
	\vspace{-2mm}
		\caption{
The nucleotide state dependence of the frequency of the burst of $\alpha 4$ helix.
This figure shows the effect of the temperature ($T= 1.25, 1.3, 1.35$) on the frequency of the burst.
\label{fig:jfig_temp_nuc_dep}}
\end{figure}

The nucleotide state dependence of the frequency of the burst correlates positively with the binding strength regardless of temperature ($T=1.25, 1.3, 1.35 < T_f$) as Fig.~\ref{fig:jfig_temp_nuc_dep}, while difference of the burst frequency  between the weak binding states (``ADP$\cdot$Pi 2'' and ``ADP'') and the strong binding states (``ATP 1'', ``ATP 2'', and ``ADP$\cdot$Pi 1'') becomes more appreciable with temperature.
For the temperature lower than $T = 1.2$, the burst is hardly observed regardless of the nucleotide states (data not shown).
Thus, at the temperature not too lower than $T_f$ the burst frequency is considerably different between the weak binding state and the strong binding state.

\subsection{The influence of the nucleotide molecule}

A protein like KIF1A work as a molecular machine only by binding the nucleotide molecule.
To investigate the influence of the presence of the nucleotide to the structural fluctuation of the whole KIF1A protein, we also simulated the system which includes both KIF1A and nucleotide molecule explicitly.
We employed the coarse-grained nucleotide model, which was introduced by Takagi and Kikuchi~\cite{takagi2007}.
Here we show only a preliminary result.
 
\begin{figure}[t]
	\begin{center}
	\includegraphics[width=7.0cm]{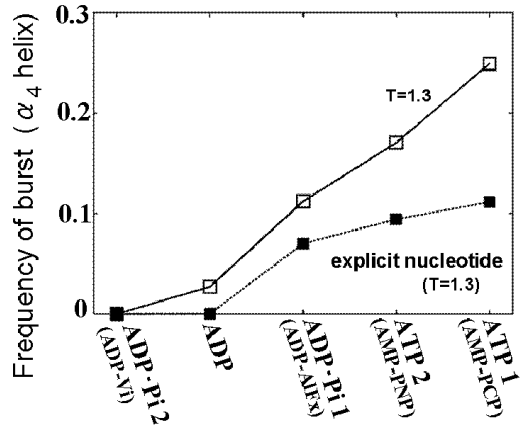}
	\end{center}
	\vspace{-2mm}
		\caption{
This figure is the intermediate state dependence of the burst frequency ($\alpha 4$ helix) at two conditions (filled square: simulation with explicit nucleotide molecule, open square: simulation without explicit nucleotide).
This figure shows the effect that the presence of nucleotide gives the frequency of the burst.
\label{fig:jfig_nuc_dep}}
\end{figure}

In Fig.~\ref{fig:jfig_nuc_dep}, we can see the same tendency of the burst frequency as before;
The burst frequency increases with the binding strength.
The effect of the explicit nucleotide is to suppress the burst frequency, in other word, the flexibility of $\alpha 4$ helix, to some extent.
More detailed simulation with the explicit nucleotide model will be left for future work.

\section{Discussion}
We have shown that according to the nucleotide states KIF1A changes the dynamics and the flexibility of the ``$\alpha 4$ helix'', which is a part of the MT binding site.
In particular, we found that the frequency of the burst of $\alpha 4$ correlated strongly with the equilibrium dissociation constants $K_{d}$ that was obtained experimentally.
This means that in the strong binding mode the MT binding site becomes more flexible than the weak binding mode.
According to these results, we suggest the possibility that $\alpha 4$ helix regulates the affinity to MT, although so far only two loops (L11, L12) were considered to be important in switching the strength of binding \cite{okada2000, kikkawa2004}.
The result also implies that the binding strength is regulated through the flexibility of the binding site.
It is consistent with the recent simulation result for the conventional Kinesin by Kenzaki and Kikuchi~\cite{kenzaki2007}, in which the ligand binding sites exhibit large structural fluctuations.
We consider that the essence of above mechanism on regulating the binding strength for the rail protein also apply to  other rail-motor systems, such as actomyosine and dynein, because the key motor elements that exhibits structural change in different nucleotide states include switch I and II commonly, the motifs that are structurally homologous to myosin and G protein regions that move upon nucleotide hydrolysis and exchange~\cite{vale2000, kull2002}.

By recent experiments~\cite{okada2000}, the equilibrium dissociation constant $K_d$ of KIF1A in the nucleotide free state is shown to be the same order as $K_d$ in the ATP bound state, although the nucleotide free structure of KIF1A as well as conventional kinesin has not been solved.
The present result suggests that judging from the value of $K_d$ the MT binding site ($\alpha 4$) in the nucleotide free structure of KIF1A should become considerably flexible.
It seems to be consistent with a recent experiment for Kar3 (Kinesin-14), which belongs to the same superfamily as KIF1A, by cryomicroscopy~\cite{hirose2006}.
The experiment revealed that the switch II helix $\alpha 4$ ``melts'' in the nucleotide free state~\cite{hirose2006}.

In Fig.~\ref{fig:jfig_histo_kd} (b), we see that the frequency of the burst correlates strongly with the equilibrium dissociation constant of CK1 (a mutant KIF1A) compared with that of CK6 (wild type KIF1A).
Since CK1 lacks a part of L12 (K-loop), which has the positive charge and interact with MT strongly, we consider that the simulated result by our model in which the charge of the amino acid residues and the interaction with MT are not taken into account corresponds to a mutant CK1 rather than CK6.
We also find that our result of burst frequency correlates strongly with the equilibrium constant $K_d$ for conventional kinesin (KK1), which also lacks K-loop (see figure 2A in ref.~\cite{okada2000}).

We have also investigated the temperature dependence of the dynamics of $\alpha 4$ helix, and found that the burst frequency of the strong binding states and the weak binding states becomes appreciably different at temperature not too lower than the folding temperature.
Thus, we suggest that the thermal fluctuation is important for switching the strength of binding on MT.

\begin{figure}[t]
	\begin{center}
	\includegraphics[width=7.0cm]{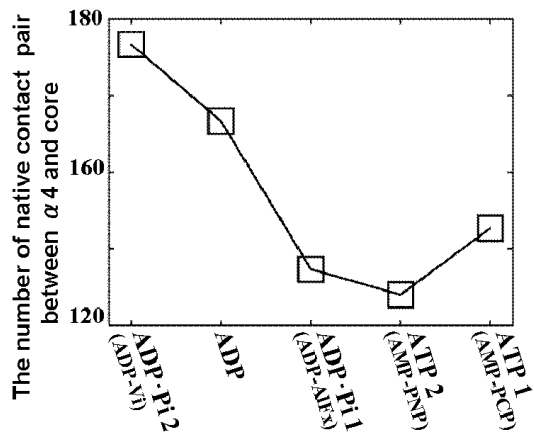}
	\end{center}
	\vspace{-2mm}
		\caption{
The intermediate state dependence of the number of native contact pair between $\alpha 4$ and the other part (head core).
\label{fig:jfig_reason}}
\end{figure}

In Fig.~\ref{fig:jfig_reason}, we show the number of native contact pair between $\alpha 4$ and core (other parts) in each intermediate state.
This figure shows that the number of the native contact pair  in the strong binding states (``ATP 1'', ``ATP 2'', and  ``ADP$\cdot$Pi 1'') is considerably smaller than that in the weak binding states (``ADP$\cdot$Pi 2'' and ``ADP'').
Thus we consider that the nucleotide state dependence of the burst frequency is a consequence of the nucleotide sate dependence of the number of the native contact pair between $\alpha 4$ and core. 

We also found that the presence of nucleotide molecule suppresses the burst.
Thus, the presence or absence of the nucleotide molecule affect the fluctuation of MT binding site which is located distant from the nucleotide binding site allosterically.

\section{Model \& method}
\subsection{G\={o}-like model}
To study large fluctuation of KIF1A in different nucleotide states we use "C$_{\alpha}$ G\={o}-like model" the version of Clementi et al.~\cite{clementi2000}, where a protein chain consists of spherical beads that represents $\rm{C}_{\alpha}$ atoms of amino acids residues connected by virtual bonds, and interactions are specified so that structures closer to the native structure are more stable.
Explicitly, the effective energy $V_{\rm p}$, at a protein conformation $\Gamma$ is given as 
\begin{eqnarray}
V_{\rm p}(\Gamma, \Gamma^{(0)}) &=& \sum_{\rm{bonds}} K_r (b_i - b^{(0)}_i)^2
		        + \sum_{\rm{angles}} K_{\theta} (\theta_i -
			\theta^{(0)}_i)^2  \nonumber \\
&+& \sum_{\rm{dihedrals}} K_{\phi}
\bigl[ 
 (1-\cos (\phi_i - \phi^{(0)}_1)) \nonumber \\
&+& \frac{1}{2} (1-\cos 3 (\phi_i - \phi^{(0)}_1)) 
\bigr] \nonumber \\
&+& \sum_{i < j-3}^{\rm{native}\ \rm{contact}} k_{\rm nc} \bigl[
5 (\frac{r_{ij}^{(0)}}{r_{ij}})^{12} -6 (\frac{r_{ij}^{(0)}}{r_{ij}})^{10} 
\bigr]	  \nonumber \\
&+& \sum_{i < j-3}^{\rm{non-native}\ \rm{contact}} k_{\rm nnc} (\frac{C}{r_{ij}})^{12}, \label{eq:Go_potential}
\end{eqnarray}
where $\Gamma^{(0)}$ signifies the native (reference) structure of protein.
The vector ${\bf r}_{ij} = {\bf r}_{i} - {\bf r}_{j}$ is the distance between the $i$th and $j$th of C$_{\alpha}$, where ${\bf r}_{i}$ is the position of the $i$th C$_{\alpha}$.
$b_i = |\bf{ b}_\textit{i}| = |\bf{ r}_{\textit{i i}+1}|$ is the virtual bond length between two adjacent C$_{\alpha}$.
$\theta_i$ is the $i$th angle between two adjacent virtual bonds, where $\cos \theta_i = ({\bf b}_{i-1}\cdot {\bf b}_{i})/(b_{i-1} b_i)$, and $\phi_i$ is the $i$th dihedral angle around ${\bf b}_{i}$. 
The first three terms of Eq.~\ref{eq:Go_potential} provide local interactions, that is, bond length, bond angle, and dihedral angle interactions, respectively.
On the other hand, the last two terms are interactions between non-local pairs that are distant along the chain.
Native contact in the fourth term is defined as follows: if one of the nonhydrogen   atoms in the $i$th amino acid is within a distance of $6.5\,\rm{\AA}$ from any nonhydrogen atom in the $j$th amino acid, we define the pair of the $i$th and $j$th amino acids as being native contact. 
Parameters with the subscript $(0)$ are the constants, of which values are taken from the corresponding variables in the native structure.
Therefore, all of the terms except the last one are set up so that each term has the lowest energy when the conformation $\Gamma$ coincides with the native structure $\Gamma^{(0)}$: this effect realizes the funnel-like energy landscape.

According to nucleotide states, KIF1A has different gap regions (missing residues) which exists mainly around two loop (L11, L12) region (L11 around 260, L12 around 290) and C terminal (around 360)~\cite{kikkawa2004, kikkawa2001}.
The length of gap regions changes according to nucleotide state.
Since in the gap regions there are not any information about native structure, we set the bond length $b^{(0)}_{i}$ in gap region $3.8\,\rm \AA$ which is the average value of protein, and set the interaction parameters in gap region ($K_{\theta}, K_{\phi}, k_{\rm nc}$) zero, whereas the values of the others (parameters $K_{r}, k_{\rm nnc}$) are hold.
Because the residues in gap regions are assumed to fluctuate freely except restraint on the bond length, the above parameter sets for gap regions may be appropriate.

If the sets for gap regions is excluded, the interactions parameter we use throughout the present work $K_{r}=100.0, K_{\theta}=20.0, K_{\phi}=1.0, k_{\rm nc}= k_{\rm nnc} = 0.25$ and $C=4.0\,\rm\AA$ are the same values as those used in Takagi and Kikuchi~\cite{takagi2007}.
(The cutoff length for calculating the forth term in Eq. \ref{eq:Go_potential} is also taken to be $2 r^{(0)}_{ij}$ like a past work~\cite{takagi2007}.)\\

\subsection{Dynamics}
The dynamics of protein are simulated by the underdamped Langevin equation at a constant temperature $T$ (in the thermal equilibrium).
\begin{align}
m_{i} {\bf \dot{v}}_{i} = {\bf F}_{i} - \gamma_i {\bf v}_{i} + {\bf \xi}_i, \label{eq:lange}
\end{align}
where ${\bf v}_{i}$ is the velocity of the $i$th bead and a dot represents the derivative with respect to time $t$ (thus, ${\bf v}_{i} = {\bf \dot{r}}_{i}$). 
${\bf F}_{i}$ and ${\bf \xi}_{i}$ are systematic and random forces on $i$th bead, respectively.
The systematic force ${\bf F}_i$ is derived from the effective energy $V_{\rm p}$ and can be defined as ${\bf F}_{i} = -\partial V_{\rm p}/\partial {\bf r}_{i}$.
${\bf \xi}_{i}$ is a Gaussian white random forces, which satisfies $\langle {\bf \xi}_{i} \rangle = 0$ and $\langle {\bf \xi}_{i}(t) {\bf \xi}_{j}(t^{'}) \rangle = 2 \gamma T \delta_{ij} \delta(t -t^{'}) {\bf 1}$, where the bracket denotes the ensemble average and ${\bf 1}$ is a $3 \times 3$ unit matrix.
Here, we note that the same unit is used both for energy and temperature and thus the Boltzmann constant $k_{B}=1$.
For a numerical integration of the Langevin equation, we use an algorithm by Honeycutt and Thirumalai~\cite{honeycutt1992}. 
We use $\gamma = 0.25$, $m_i$=1.0, and the finite time step $\Delta t =0.02$.

For a given protein conformation $\Gamma$, we define that the native contact between $i$ and $j$ is formed if the distance $r_{ij}$ is $< 1.2 r_{ij}^{(0)}$, where $r^{(0)}_{ij}$ is the distance between the $i$th and $j$th amino acids at the native structure $\Gamma^{(0)}$.
We then use a standard measure of the nativeness, $Q(\Gamma)$, for a given protein conformation $\Gamma$, defined as the ratio of numbers of formed native contacts at $\Gamma$ to those at the native structure $\Gamma^{(0)}$.

\begin{acknowledgments}
We thank Shoji Takada and Kazuo Sasaki for valuable discussions.
\end{acknowledgments}


\begin{thebibliography}{24}

\bibitem{howard2001} Howard, J. (2001) in {\it Mechanics of Motor
Proteins and the Cytoskeleton}(Sinauer Associates), pp. 229-244 and 263-283.

\bibitem{alberts2002} Alberts, B., Johnson, A., Lewis, J., Raff, M., 
Roberts, K., and Walter, P. (2002) in {\it Molecular Biology of
the Cell} (Garland Science), pp. 949-969.

\bibitem{kikkawa2004} Nitta, R., Kikkawa, M., Okada, Y., and Hirokawa,
N. (2004) {\it Science} {\bf 305}, 678-683.

\bibitem{okada2000} Okada, Y. and  Hirokawa, N. (2000) {\it
Proc. Natl. Acad. Sci. USA} {\bf 97}, 640-645.

\bibitem{okada1999} Okada, Y. and  Hirokawa, N. (1999) {\it Science} {\bf 283}, 1152-1157.

\bibitem{okada2003} Okada, Y., Higuchi, H., and Hirokawa, N. (2003) {\it Nature} {\bf 424}, 574-577.

\bibitem{kikkawa2001} Kikkawa, M.,  Sablin, E. P., Okada, Y., Yajima, H., Fletterick, R. J., and Hirokawa, N. (2001) {\it Nature} {\bf 411}, 439-445.

\bibitem{kikkawa2000} Kikkawa, M.,  Okada, Y., and Hirokawa, N. (2000) {\it Cell} {\bf 100}, 241-252.

\bibitem{sack1999} Sack, S.,  Kull, F. J., and Mandelkow, E. (1999) {\it Eur J Biochem} {\bf 262}, 1-11.

\bibitem{zheng2003} Zheng, W. and Doniach, S. (2003) {\it Proc. Natl. Acad. Sci. USA} {\bf 100}, 13253-13258.

\bibitem{kenzaki2007} Kenzaki, H. and Kikuchi, M. (2007) {\it Proteins} DOI: 10.1002/prot.21707.

\bibitem{brygenlson1995} Bryngelson, J. D., Onuchic, J. N., Socci, N. D., and Wolynes, P. G. (1995) {\it Proteins} {\bf 21}, 167-195.

\bibitem{onuchic1997} Onuchic, J.N., Luthey-Schulten, Z., and Wolynes, P. G. (1997) {\it Annu Rev Phys Chem} {\bf 48}, 545-600.

\bibitem{go1983} Go, N. (1983) {\it Annu. Rev. Biophys. Bioeng.} {\bf 12}, 183-210.

\bibitem{koga2006} Koga, N. and Takada, S. (2006) {\it Proc. Natl. Acad. Sci. USA.} {\bf 103}, 5367-5372.

\bibitem{hyeon2007} Hyeon, C. and Onuchic, J. N. (2007) {\it Proc. Natl. Acad. Sci. USA.} {\bf 104}, 2175-2180.

\bibitem{hyeon2007_2} Hyeon, C. and Onuchic, J. N. (2007) {\it Proc. Natl. Acad. Sci. USA.} {\bf 104}, 17382-17387.
 
\bibitem{clementi2000} Clementi, C., Nymeyer, H., and Onuchic, J. N. (2000) {\it J. Mol. Biol.} {\bf 298}, 937-953.

\bibitem{takagi2007} Takagi, F. and Kikuchi, M. (2007) {\it Biophys. J.} {\bf 93}, 3820-3827.

\bibitem{honeycutt1992} Honeycutt, J. D. and Thirumalai, D. (1992) {\it Biopolymers} {\bf 32}, 695-709.

\bibitem{vale2000} Vale, R. D. and Milligan, R. A. (2000) {\it Science} {\bf 288}, 88-95.

\bibitem{kull2002} Kull, F. J. and Endow, S. A. (2002) {\it J. Cell Sci.} {\bf 115}, 15-23.

\bibitem{amos2007} Amos, L. A. and Hirose, K. (2007) {\it J. Cell Sci.} {\bf 120}, 3919-3927.

\bibitem{hirose2006} Hirose, K., Akimaru, K. E., Akiba, T., Endow, S. A., and Amos, L. A. (2006) {\it Mol. Cell.} {\bf 23}, 913-923.

\end{thebibliography}
\end{document}